\documentclass[5p,sort&compress]{elsarticle}
\usepackage{graphicx}
\usepackage{latexsym}
\usepackage{wrapfig}
\usepackage{epsfig,color,rotating,amsmath,delarray,array}
\usepackage{pifont,float,amssymb}
\usepackage[normalem]{ulem}
\usepackage[caption=false]{subfig}
\usepackage{fixltx2e}
\usepackage{setspace}
\usepackage{placeins}
\definecolor{gray01}{gray}{0.9}
\definecolor{gray02}{gray}{0.8}
\definecolor{gray03}{gray}{0.7}
\definecolor{gray04}{gray}{0.6}
\definecolor{gray05}{gray}{0.5}
\definecolor{gray06}{gray}{0.4}
\definecolor{gray07}{gray}{0.3}
\definecolor{gray08}{gray}{0.2}
\definecolor{gray09}{gray}{0.1}
\definecolor{cyanaw}{rgb}{0,1,1}
\definecolor{lepsaw}{RGB}{229,95,254}
\definecolor{magentaaw}{rgb}{1,0,1}

\newcommand{\be}{\begin{eqnarray}}
\newcommand{\ee}{\end{eqnarray}}

\newcommand{\bgpom}{\mbox{\color{red}\rule[0.15ex]{0.25cm}{2pt}\hspace{1.5mm}\rule[0.15ex]{0.04cm}{2pt}\hspace{1.5mm}\rule[0.15ex]{0.25cm}{2pt}}}
\newcommand{\bgoneminus}{\mbox{\color{green}\rule[0.05ex]{0.085cm}{2pt}\hspace{0.65mm}\rule[0.05ex]{0.03cm}{2pt}\hspace{0.65mm}\rule[0.05ex]{0.085cm}{2pt}}}
\newcommand{\bgthreeplus}{\mbox{\color{blue}\rule[0.15ex]{0.2cm}{2pt}\hspace{0.5mm}\rule[0.15ex]{0.2cm}{2pt}}}
\newcommand{\bgthreeminus}{\mbox{\color{magentaaw}\rule[0.05ex]{0.09cm}{2pt}\hspace{0.65mm}\rule[0.05ex]{0.09cm}{2pt}\hspace{0.65mm}\rule[0.05ex]{0.09cm}{2pt}}}
\newcommand{\bgfiveplus}{\mbox{\color{cyanaw}\rule[0.05ex]{0.12cm}{2pt}\hspace{0.5mm}\rule[0.05ex]{0.025cm}{2pt}\hspace{0.5mm}\rule[0.05ex]{0.025cm}{2pt}\hspace{0.5mm}\rule[0.05ex]{0.12cm}{2pt}}}

\newcommand{\onehalfminus}{$1/2^-$}
\newcommand{\threehalfplus}{$3/2^+$}
\newcommand{\threehalfminus}{$3/2^-$}
\newcommand{\fivehalfplus}{$5/2^+$}

\begin{document}

\title{Photoproduction of $\omega$ Mesons off the Proton\tnoteref{t1}}
\tnotetext[t1]{CBELSA/TAPS Collaboration}

\address[HISKP]{Helmholtz-Institut f\"ur Strahlen- und Kernphysik, Universit\"at Bonn, D-53115 Bonn, Germany}
\address[FSU]{Department of Physics, Florida State University, Tallahassee, Florida 32306, USA}

\address[GATCHINA]{Petersburg Nuclear Physics Institute, RU-188350 Gatchina, Russia}
\address[PI]{Physikalisches Institut, Universit\"at Bonn, D-53115 Bonn, Germany}
\address[KVI]{KVI, 9747 AA Groningen, The Netherlands}
\address[GIESSEN]{II. Physikalisches Institut, Universit\"at Gie\ss en, D-35392 Gie\ss en, Germany}
\address[BASEL]{Physikalisches Institut, Universit\"at Basel, CH-4056 Basel, Switzerland}
\address[JUELICH]{Institut f\"ur Kernphysik, Forschungszentrum J\"ulich, D-52428 J\"ulich, Germany}
\address[BOCHUM]{Institut f\"ur Experimentalphysik I, Ruhr-Universit\"at Bochum, D-44780 Bochum, Germany}
\cortext[cor1]{Corresponding author}

\author[HISKP,FSU]{A.~Wilson}
\author[FSU]{V.~Crede\corref{cor1}} 
\ead{crede@fsu.edu}

\author[HISKP,GATCHINA]{A.V.~Anisovich} 
\author[KVI]{J.C.S. Bacelar}
\author[PI]{B.~Bantes} 
\author[HISKP]{O.~Bartholomy} 
\author[HISKP,GATCHINA]{D.~Bayadilov}
\author[HISKP]{R.~Beck}
\author[GATCHINA]{Y.A.~Beloglazov}
\author[GIESSEN]{K.T.~Brinkmann} 
\author[KVI]{R.~Castelijns}
\author[PI]{H.~Dutz}
\author[PI]{D.~Elsner} 
\author[PI]{R.~Ewald} 
\author[PI]{F.~Frommberger}
\author[HISKP]{M.~Fuchs} 
\author[HISKP]{Chr.~Funke}
 \author [GIESSEN]{R.~Gregor}
\author[GATCHINA]{A.~Gridnev}
\author[GIESSEN,HISKP]{E.~Gutz} 
\author[PI]{J.~Hannappel} 
\author[PI]{W.~Hillert} 
\author[HISKP]{P.~Hoffmeister} 
\author[HISKP]{I.~Horn} 
\author[BASEL]{I.~Jaegle} 
\author[PI]{T. Jude}
\author[HISKP]{J.~Junkersfeld} 
\author[HISKP]{H.~Kalinowsky} 
\author[PI]{V.~Kleber} 
\author[PI]{Frank~Klein} 
\author[PI]{Friedrich~Klein} 
\author[HISKP]{E.~Klempt} 
 \author[BASEL,GIESSEN]{M.~Kotulla} 
\author[BASEL]{B.~Krusche} 
\author[HISKP]{M.~Lang} 
\author[KVI]{H.~L\"ohner} 
\author[GATCHINA]{I.V.~Lopatin} 
 \author[GIESSEN]{S.~Lugert} 
\author[BASEL]{T.~Mertens} 
\author[KVI,GIESSEN]{J.G.~Messchendorp} 
\author[GIESSEN]{V.~Metag} 
\author[GIESSEN]{M.~Nanova} 
\author[HISKP,GATCHINA]{V.A.~Nikonov} 
\author[GATCHINA]{D.~Novinski} 
\author[GIESSEN]{R. Novotny} 
\author[PI]{M.~Ostrick} 
\author[GIESSEN]{L.M. Pant} 
\author[HISKP]{H.~van~Pee} 
\author[GIESSEN]{M.~Pfeiffer} 
\author[GIESSEN]{A.~Roy} 
\author[HISKP,GATCHINA]{A.V.~Sarantsev} 
\author[HISKP]{C.~Schmidt} 
\author[PI]{H.~Schmieden} 
\author[KVI]{S.~Shende} 
\author[HISKP]{V.~Sokhoyan} 
\author[FSU]{N.~Sparks}
\author[PI]{A.~S{\"u}le} 
\author[GATCHINA]{V.V.~Sumachev} 
\author[HISKP]{T.~Szczepanek} 
\author[HISKP]{U.~Thoma} 
\author[GIESSEN]{D.~Trnka} 
\author[GIESSEN]{R.~Varma} 
\author[HISKP,PI]{D.~Walther} 
\author[HISKP]{Ch.~Wendel} 
\author[BOCHUM]{U.~Wiedner} 

\begin{abstract}
The differential cross sections and unpolarized spin-density matrix elements for the reaction 
$\gamma p\to p\omega$ were   measured using the CBELSA/TAPS experiment for initial photon energies
ranging from the reaction threshold to 2.5 GeV. These observables were measured from the radiative decay 
of the $\omega$ meson, $\omega\to\pi^0\gamma$. The cross sections cover the full angular range and show 
the full extent of the $t$-channel forward rise. The overall shape of the angular distributions in the differential 
cross sections and unpolarized spin-density matrix elements are in fair agreement with previous data. In 
addition, for the first time, a beam of linearly-polarized tagged photons in the energy range from 1150~MeV 
to 1650~MeV was used to extract polarized spin-density matrix elements.   

These data were included in the Bonn-Gatchina partial wave analysis (PWA). The dominant contribution to 
$\omega$~photoproduction near threshold was found to be the \threehalfplus{} partial wave, which is primarily 
due to the sub-threshold $N(1720)$\,\threehalfplus{} resonance. At higher energies, pomeron-exchange was found to 
dominate whereas $\pi$-exchange remained small. These $t$-channel contributions as well as further contributions 
from nucleon resonances were necessary to describe the entire dataset: the \onehalfminus{}, \threehalfminus{}, and 
\fivehalfplus{} partial waves were also found to contribute significantly. 
\end{abstract}

\date{Received: \today / Revised version:}

\maketitle

\section{Introduction}
\label{sec:Intro}
The spectrum of excited states has historically given essential information on the nature of any 
composite quantum system. The careful mapping of the excited states of baryons shines light on the 
nature of the nonperturbative regime of quantum chromodynamics (QCD). This spectrum specifically 
depends on the effective degrees of freedom and the forces confining the quarks. Symmetric quark models, 
which attempt to describe the baryon system, predict the pattern of low-mass baryons reasonably well. 
However, the predicted baryon states for masses above 1.8~GeV/$c^2$ greatly outnumber those which have
been found experimentally. Most known light-flavor baryon resonances lie below 2 GeV/$c^2$ and were
discovered in elastic $\pi N$ scattering experiments. Quark model calculations have shown that many of 
these so-called ``missing'' baryons have weak $\pi N$ couplings; and moreover, they could strongly couple 
to $\eta N$ and $\omega N$ without a small coupling to $\gamma N$. In recent years, many laboratories 
around the world (ELSA, GRAAL, Jefferson Laboratory,  MAMI,  SPring-8, etc.) have published differential cross 
sections and polarization observables in photoproduced reactions. For a recent review on baryon resonances, 
see \cite{Klempt:2009pi,Crede:2013kia}. 

According to the predictions of the Constituent Quark Model, e.g.~\cite{Capstick:1993kb}, data on 
$\omega$ photoproduction have a good chance of revealing some of the ``missing'' baryon resonances.  
However, since the $\omega$~meson has the same quantum numbers as the incoming photon, meson 
exchange ($t$-channel) processes are likely to contribute strongly. To disentangle the $t$-channel from 
the resonant ($s$-channel) amplitude, data with full angular coverage are needed. Of particular importance 
is the very forward direction where the $t$-channel amplitude has its maximum. 

Moreover, the $\omega$ meson acts as an isospin filter for baryon resonances. Since the isospin of the 
$\omega$ meson is zero, any baryon resonance decaying to $N\omega$ must have isospin $I = 1/2$, and 
therefore contributions from $\Delta$ states are excluded.  

In this paper, the differential cross sections and spin-density matrix elements for the reaction 
\begin{equation} \gamma p \to p\omega\label{eqn:reaction}\end{equation} 
are presented by reconstructing the $\omega$ from the neutral decay,
\begin{equation} \omega \to \pi^0 \gamma \to \gamma\gamma\gamma\:.\end{equation} 

\section{Experimental Setup}
\label{sec:ExpSetup}
The CBELSA/TAPS experiment was conducted at the electron stretcher accelerator (ELSA) facility~\cite{ELSApast} 
located at the University of Bonn in Germany. A 3.175~GeV electron beam from ELSA interacted with a radiator 
target and produced bremsstrahlung photons. The radiator target was situated in a goniometer which contained 
copper radiators of varying thickness along with a diamond radiator for linear polarization. The unpolarized data 
utilized a copper radiator of thickness 3/1000\,$X_R$ (radiation length). The polarized data used a diamond 
radiator. The bremsstrahlung electrons were deflected by a dipole magnet into the tagging detector system 
(tagger). The tagger consisted of 480 scintillating fibers on top of 14~scintillating bar counters which partly 
overlapped. Using the knowledge of the magnetic field strength and the hit position in the tagger, the energy 
of each electron was determined and used to tag each bremstrahlung photon with energy and time information. 

A fraction of the tagged bremsstrahlung photons continued down the beam line and interacted with the 
protons in the liquid hydrogen target to produce mesons which decayed to final-state photons. The energy 
and position of these photons were detected by the crystal modules in the two electromagnetic calorimeters, 
Crystal Barrel and TAPS. The Crystal Barrel detector, in its configuration during the CBELSA/TAPS experiment of 
2002/2003, consisted of 1290 CsI(Tl) crystals, which were read out by photodiodes. The TAPS detector consisted 
of 528~BaF$_2$ crystals, which were read out by photomultiplier tubes. They formed a hexagonal wall that covered 
the forward hole left open by the Crystal Barrel detector. Together the Crystal Barrel and TAPS detectors covered 
more than 98\,\% of the full $4\pi$ solid angle. Protons or any other charged particles were identified by either 
5~mm thick plastic scintillators placed in front of each TAPS crystal or by a three layer scintillating fiber detector 
which closely surrounded the target. For more information on this setup, see~\cite{CREDEeta}.

\section{Data Analysis}
\label{sec:Danalysis}
The unpolarized data were recorded in October 2002 and November 2002. The linearly-polarized data 
were recorded in March and May 2003. The polarized data used a diamond radiator optimized to have a 
coherent polarization edge at 1350~MeV and 1600~MeV with a polarization maximum of 49\% and 39\%,
respectively. More information on the goniometer and the linear beam polarization can be found in 
\cite{Elsner:2008sn}. The ELSA beam energy for the unpolarized runs was 3.175~GeV, but for this analysis, 
only photons up to 2.55~GeV were used due to the lack of tagger scintillating fibers above this energy. The 
fibers provided a fine energy resolution and additional timing information. The trigger for these datasets 
relied on Leading-Edge Discriminator~(LED) outputs which signaled if the energy deposit in a group of TAPS 
crystals was above either a low-energy threshold~(LED-low) or a higher-energy threshold~(LED-high). Each 
TAPS crystal belonged to one of eight LED-low sectors and one of eight LED-high sectors. The trigger required 
either 1) two LED-low sectors in the TAPS detector firing above a low energy threshold or 2) one LED-high sector
in TAPS above a higher energy threshold and at least one hit (two hits for the polarized data) in the Crystal Barrel. 
For sector definitions and more information, see~\cite{CREDEeta,Sparks:2010vb}. The same data were used for 
several previously published analyses on a variety of final
states~\cite{CREDEeta,Elsner:2008sn,Credepi0,NanovaKstarSigma,Ewald:2011gw,Castelijns:2007qt,Gutz:2009zh,
Sparks:2010vb,Klein:2008aa,Elsner:2007hm}. 

In order to study Reaction~(\ref{eqn:reaction}), the $p \pi^0 \gamma$ final state was reconstructed first 
using kinematic fitting. All events based on three distinct neutral hits and less than two charged hits were 
subjected to the $\gamma p\to p_{\rm missing}\,\pi^0\gamma$ hypothesis. The $\omega$ yields were then 
extracted from $\pi^0\gamma$ invariant masses by carefully subtracting the background contribution. The 
proton was chosen as a missing particle in the fit due to the relatively large uncertainty in reconstructing the 
proton energy and momentum from calorimeter output. The resulting confidence-level (CL) values from kinematically 
fitting the data events and Monte Carlo simulated $p\omega$ events were used first to reduce background in the 
analysis. A cut of CL$_{p_{\rm missing}\,\pi^0\gamma}\:>\:0.005 \label{eqn:CLcut}$ was applied. This very small CL 
cut simply guaranteed convergence of the kinematic fit, and therefore energy-momentum conservation, but 
had essentially no impact on the $p\omega$~yield. The remaining $\omega$~background events were removed 
by applying a probabilistic method which is described below. More information on kinematic fitting used at the 
CBELSA/TAPS experiment can be found in~\cite{CREDEeta}.

To isolate the incoming photon, a coincident timing cut between the tagger and TAPS was used to reduce 
the number of initial photon candidates. The remaining photons were subjected to kinematic fitting, which
required energy and momentum conservation. The beam photon with the largest CL value was chosen as the 
initial photon. An equivalent analysis of events with timing outside of this coincident timing cut was performed 
and used to eliminate the effect of accidental background.

The contribution of $p\pi^0$ events which were poorly reconstructed as $p\pi^0\gamma$~events could 
be effectively separated from good $p\omega$~events by studying the momentum-depen\-dence of the 
opening angle between the $\pi^0$ and the final-state bachelor-photon in the center-of-mass frame, 
$\theta_{c.m.}^{\pi^0,\gamma}$. In this two-dimensional distribution, events which satisfied 
$|p^{\omega}_{c.m.}| < (-13.33\; \theta_{c.m.}^{\pi^0,\gamma} + 2400)\;\text{MeV}/c \label{eqn:OpAC}$ were 
kept for further analysis. Simulations showed that this cut removed $p\pi^0$ events which caused 
structures in the $\pi^0\gamma$ background mass and facilitated the modeling of the background.

When a proton interacted with TAPS, the energy deposit was typically smaller than for a photon of the same 
energy. The TAPS crystal trigger thresholds, LED-low and LED-high, were calibrated for photon triggering and 
responded much differently for protons. Due to this issue, triggering on low-momentum proton energy deposits 
was difficult to model correctly and could have caused errors in the detector acceptance correction. Fortunately, 
the kinematics of the reaction allowed this class of events to be excluded from the analysis without losing angular 
coverage. All events which had an incident photon energy less than 1600~MeV, a proton reconstructed in TAPS, 
and no additional photons triggering in TAPS above the LED-high threshold were cut out of both the data and 
the Monte Carlo events. The initial photon energy requirement was chosen to ensure that only low-momentum 
protons were discarded. The detector response to high-energy protons was similar to that of photons.
For more details on the $\omega$~event selection, please see~\cite{DissWilson}.

\begin{figure}[bt]
 \centering
\label{fig:Total_inv_mass_data}\includegraphics[width=0.49\textwidth]{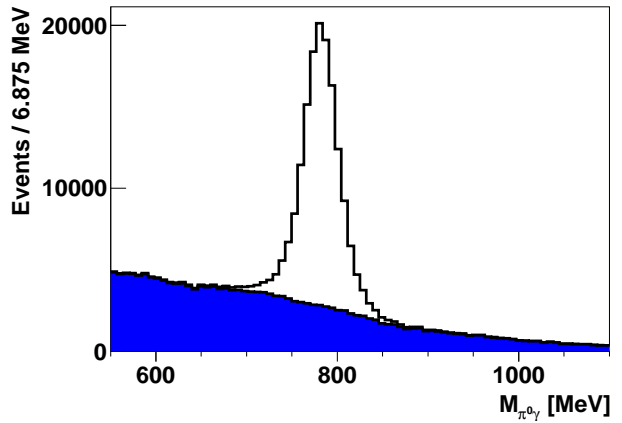}
\caption[Final $\pi^0\gamma$ Invariant Mass distribution]{(Color Online) The invariant $\pi^0\gamma$ 
  mass distribution for unpolarized data events which were subjected to the Q-factor fitting (background 
  subtraction). These events survived all kinematic cuts. The solid blue area shows the background subtracted 
  using the Q-factor method. There were approximately 128,000 $\omega$'s in this dataset.}
\label{fig:Total_inv_mass}
\end{figure}

To further remove non-$p\omega$ events (background) from the data, a method was applied which assigned 
a quality factor (Q-factor) to each event. This factor describes the probability for an event to be a $p\omega$ event. 
The Q-factors were then used to weight each event in the analysis when an observable was formed. This method
is described in more detail in~\cite{Williams:2008sh} and its previous application to the CBELSA/TAPS experiment 
in \cite{Sparks:2010vb}. Figure~\ref{fig:Total_inv_mass} shows the resulting separation of signal and background
in the total $\pi^0\gamma$ invariant mass distribution. The number of experimentally observed $p \omega$ data 
events was approximately 128,000 in the unpolarized dataset and 60,000 in the polarized combined datasets. 

The response of the CBELSA/TAPS experiment was studied with a GEANT3-based Monte Carlo simulation. 
The response of the Crystal Barrel and TAPS detectors to photons was well reproduced by the Monte Carlo 
simulations. The acceptance for the $p\omega$ events was determined by simulating events which were 
evenly distributed over the available kinematic phase space. The detector acceptance is defined as the ratio 
of reconstructed Monte Carlo events to generated Monte Carlo events for each kinematic bin. The tagging 
and timing of initial state photons were not simulated. The Monte Carlo events were subject to exactly the 
same reconstruction as the data events from the experiment. The Q-factor method of background
subtraction was not applied to the Monte Carlo events which are background-free by construction.

The angular distributions of the decay products of the $\omega$~meson were analyzed to obtain more 
information about the production mechanisms leading to this final state. For example, the spin-density 
matrix elements (SDMEs) are a frame-dependent expression of the $\omega$~meson helicity. 

The unpolarized SDMEs were extracted by combining the polarized and unpolarized datasets in a fit. They were 
extracted from the data for each kinematic bin by performing an Extended Maximum Likelihood fit to \cite{Zhao:2005vh} 

\begin{eqnarray} W^0(\theta_d,\phi_d,\rho^0)& =& \frac{3}{8\pi} \{ \sin^2\theta_d\;\rho^0_{00}\label{Eqn:SDME0} \\[1ex]
& & +  \frac{1}{2}\left( 1+\cos^2\theta_d \right)(1-\rho^0_{00})   \nonumber\\[1ex]
& & + \sin^2\theta_d \cos 2\phi_d\,\rho^0_{1-1} \nonumber\\[0.6ex]
 & & + \sqrt{2} \sin 2\theta_d \cos\phi_d\,\text{Re}\rho^0_{10}   \} \;,\nonumber 
\end{eqnarray} 
where $\rho^0$ is the unpolarized spin-density matrix, and $\theta_d$ and $\phi_d$ are the polar and the 
azimuthal angle of the $\omega$ bachelor-photon measured in the rest frame of the $\omega$ meson. 
Acceptance corrections were taken into account in the likelihood function during the fitting. The statistical 
uncertainties for these data were studied by using a large number of Monte Carlo datasets with varying numbers 
of events and with the same SDME values. The uncertainty in the fit parameters as a function of the number of 
events in the dataset was studied and used to define the overall statistical uncertainty for each reported SDME. 
All SDMEs in this analysis were extracted in the helicity, Gottfried-Jackson, and Adair frames~\cite{Schilling:1969um}.  
 
\begin{figure*}[t!]
 \centering
 \includegraphics[width=0.95\textwidth,height=0.35\textheight]{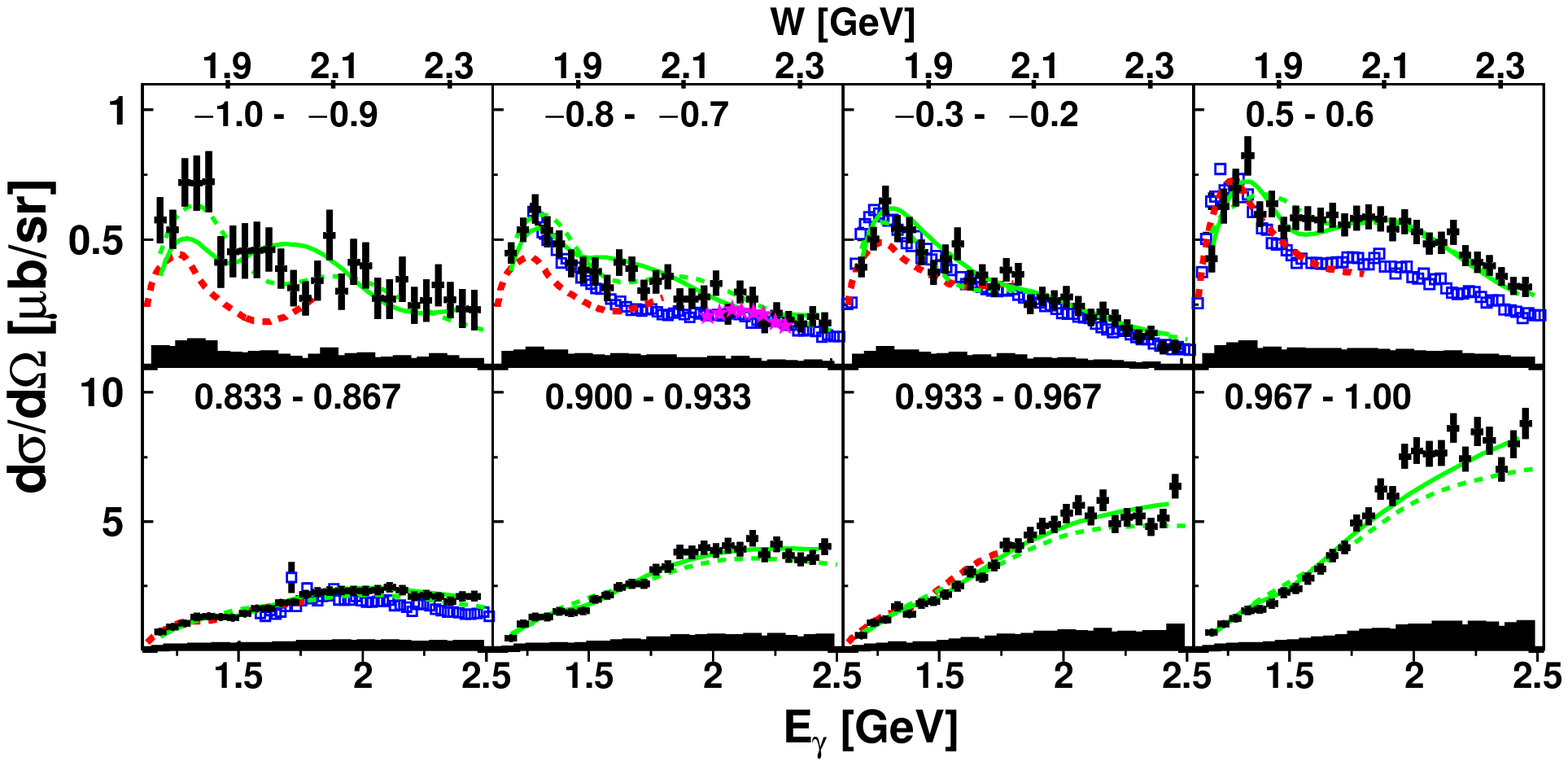}
 \caption[Excitation functions]{(Color Online) Excitation functions for $\gamma p \to p \omega$ from 
  CBELSA/TAPS ($\bullet$) for selected angle bins. For comparison, the CLAS data~\cite{Williams:2009yj} 
  are represented by {\tiny \color{blue}$\square$}, the LEPS data \cite{LEPSeta} are represented by {\scriptsize
  \color{lepsaw}$\bigstar$} (for the $-0.8$ - $-0.7$ panel), and the Gie\ss en Lagrangian fit in \cite{Shklyar:2004ba} 
  is represented by {\color{red}\rule[0.5ex]{0.1cm}{1pt} \hspace{0.02mm}\rule[0.5ex]{0.1cm}{1pt} 
  \hspace{0.02mm}\rule[0.5ex]{0.1cm}{1pt}} . The Bonn-Gatchina PWA solutions are represented by green lines:
  {\color{green}\rule[0.5ex]{0.34cm}{1pt}} (full solution), 
  {\color{green}\rule[0.5ex]{0.1cm}{1pt}\,\hspace{0.02mm}\rule[0.5ex]{0.1cm}{1pt}\,\hspace{0.02mm}\rule[0.5ex]{0.1cm}{1pt}}
  (solution without \threehalfplus{} partial wave). Statistical uncertainties are reported as vertical bars on each 
  data point. The total systematic uncertainty for the CBELSA/TAPS data is shown as a black band at the bottom 
  of each plot. Each plot is labeled with its range in cosine of the polar angle in the center-of-mass frame of the
  $\omega$ meson ($\cos \theta^{\,\omega}_{\rm c.m.}$). The horizontal axis is measured in the energy of the initial 
  photon. An additional horizontal axis at the top of the figure shows the center-of-mass energy.}
\label{fig:Dcross_EF}
\end{figure*}

Polarized SDMEs were extracted from data by combining the two polarized datasets in a fit where each 
kinematic bin was subjected to a Maximum Likelihood fit to~\cite{Zhao:2005vh,Schilling:1969um} 
\begin{eqnarray}  W^L(\Omega_d,\Phi_{\rm pol},\rho) &=& W^0(\Omega_d,\rho)\label{Eqn:SDMEPL}\\
 &&+\frac{3}{8\pi} P_\gamma\,[ \cos\Phi_{\rm pol}W_1(\Omega_d,\rho)
       \nonumber\\&&+ \sin\Phi_{\rm pol}W_2(\Omega_d,\rho) ]  \nonumber \\[2ex]
  {\rm with}~~W_1(\Omega_d,\rho) &=& \sin^2\theta_d\,\rho^1_{00} \label{Eqn:SDMEP1} \\
 &&+ (1+\cos^2\theta_d)\,\rho^1_{11} \nonumber\\&&+ \sin^2\theta_d\cos 2 \phi_d\,\rho^1_{1-1}\nonumber\\
 &&+\sqrt{2}\sin 2 \theta_d\cos \phi_d\,\text{Re}\,\rho^1_{10}\nonumber\\[1ex]
   W_2(\Omega_d,\rho) &=& \sin^2\theta_d\sin 2 \phi_d \,\text{Im}\,\rho^2_{1-1}\label{Eqn:SDMEP2}\\
 &&+ \sqrt{2}\sin 2\theta_d\sin\phi_d\,\text{Im}\,\rho^2_{10}\nonumber  \;,\end{eqnarray} 
where $P_\gamma$ is the degree of polarization of the photon beam and $\Phi_{\rm pol} $ is the polarization 
angle. Statistical uncertainties were estimated in the same manner as for the unpolarized SDMEs. The values 
of the polarized SDMEs were limited in the fit by the restrictions listed in \cite{Schilling:1969um}. 
   
For the differential cross sections, the statistical errors were determined from the number of events in each 
$(E_\gamma,\,{\rm cos}\,\theta_{\rm c.m.})$ bin. The systematic uncertainties are given as error bands at the 
bottom of each distribution in Figure~\ref{fig:Dcross_EF}; the statistical and systematic uncertainties for 
each data point in Figures~\ref{fig:SDME_unpol} and \ref{fig:SDME_pol}~(SDMEs) were added in quadrature. 
Sources of the systematic uncertainties include kinematic fitting and a possible target shift away from the 
known position in the Monte Carlo simulation. The corresponding uncertainties were determined by applying 
different CL cuts as well as varying the target position in the Monte Carlo ($\pm\,1.5$~mm~\cite{CREDEeta}) 
and evaluating changes in the re-extracted observables. The effects were observed to be $<$\,1.5\,\% and 
$<$\,4\,\% on average, respectively. The errors of the decay branching fractions were negligible. Further
contributions include Q-factor fitting errors of $<$\,8\,\% (for details on determining the methodical errors,
see~\cite{Williams:2008sh}), photon flux uncertainties of $<$\,10\,\% for the differential cross sections, and 
a 5\,\% degree of polarization uncertainty for the polarized SDMEs. All these errors were then added 
quadratically to give the total systematic error.

\section{Experimental Results}
\label{sec:EResults}
The differential cross sections, $d\sigma/d\Omega$, for $\gamma p\to p\omega$ from this analysis are 
shown in Figure~\ref{fig:Dcross_EF}~(black dots) for a few selected angle bins. These serve as representative 
distributions of the entire dataset. The data are binned in 50~MeV-wide initial photon energy bins from 1.15~GeV 
to 2.5~GeV. For $-1.0 < {\rm cos}\,\theta_{\rm c.m.}^{\,\omega} < 0.8$, the angular bin width is 0.1. To show the forward
region's ($0.8 < {\rm cos}\,\theta_{\rm c.m.}^{\,\omega} < 1.0$) strong rise, the bins are 0.033 wide.  The vertical error 
bars on each point are statistical only and the systematic uncertainties, on the order of 15\,\%, are shown as a black 
band at the bottom of each distribution. The CLAS~\cite{Williams:2009yj} and LEPS cross sections \cite{LEPSeta} are 
also shown for comparison. The dashed red line represents a calculation within a coupled-channel effective Lagrangian 
approach in the energy region from the pion threshold up to 2~GeV (Gie\ss en model)~\cite{Shklyar:2004ba}, which 
also includes data on $\pi$-induced reactions. The Gie\ss en group reported a reasonable description of the SAPHIR 
data~\cite{SAPHIRw} in their analysis~(\cite{Shklyar:2004ba},~Figure~9 therein). Since the SAPHIR data were available 
only as a function of the mandelstam variable~$t$, the Gie\ss en model is shown as a representation of that dataset. 

\begin{figure}[t]
 \centering
\includegraphics[width=0.46\textwidth]{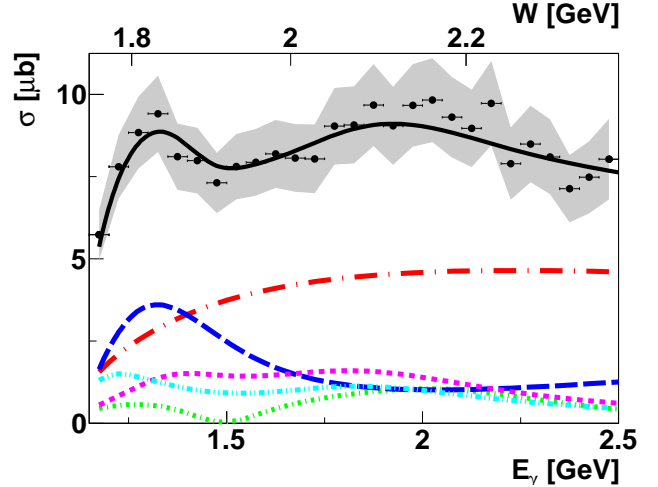}
\caption[Spin Density Matrix Total Cross Section]{(Color Online) Total cross cection for $\gamma p\to p\omega$ 
from the CBELSA/TAPS experiment ($\bullet$) as a function of the initial photon ($E_{\gamma}$) and center-of-mass (W) 
energy. The error bars represent the statistical uncertainty and the gray band represents the systematic uncertainty. The 
Bonn-Gatchina fit is represented with a solid black line. The largest contributions to this fit are (listed from largest to 
smallest at the highest energies)  pomeron-exchange~(\bgpom), resonant production of the $J^p$ = $3/2^+$ partial 
wave~(\bgthreeplus), $3/2^-$ partial wave~(\bgthreeminus), $5/2^+$ partial wave~(\bgfiveplus), and $1/2^-$ partial 
wave~(\bgoneminus).}
\label{fig:TotCross}
\end{figure}

The differential cross sections are reported over the full kinematic phase space, and they can be integrated over 
to measure the total cross section without any extrapolation. The total cross section in Figure~\ref{fig:TotCross} 
is reported for initial photon energies ranging from 1.15~GeV to 2.5~GeV. 

The overall shape of the various experimental distributions is in fair agreement. However, some systematic 
differences can be seen, in particular in the very backward direction and at higher energies. The earlier results 
from SAPHIR, as represented by the Gie\ss en model, appear to be systematically lower for 
${\rm cos}\,\theta_{\rm c.m.}^{\,\omega} < 0.0$ than all other results, whereas an almost linear energy-dependent 
normalization discrepancy is observed between CLAS and CBELSA/TAPS. It is worth noting that older SLAC 
results~\cite{Ballam:1972eq} at 2.8~GeV exhibit a similar normalization discrepancy with CLAS as do the new 
data presented here.

It is important to resolve the normalization discrepancy between the results from CBELSA/TAPS and CLAS for a 
reliable extraction of physics contributions to the cross section. A similar normalization discrepancy has been 
observed recently between the two experiments for the reaction $\gamma p\to p\eta$~\cite{CREDEeta}. The very 
similar behavior of the discrepancies in both reactions hints again at a normalization issue. Significant efforts have 
been invested in both collaborations to understand this normalization issue, but the nature of the discrepancy has
remained unclear.

The shape of the $\gamma p\to p\omega$ differential cross sections suggests two dominant processes. The 
low-energy cross sections are nearly flat and suggest $s$-channel resonance production. The effect of this 
resonance production can be seen at low energies in the total cross section as a resonant peak. In the 
higher-energy region, the increase in the cross sections toward ${\rm cos}\,\theta_{\rm c.m.}^{\,\omega}\sim1$ 
indicates that a $t$-channel or meson exchange process contributes strongly. For $W < 2$~GeV, any $u$-channel
 (baryon exchange) contribution must be small due to the lack of any visible peak in the backwards cross section.

\begin{figure*}[t!]
 \centering
 \includegraphics[width=0.89\textwidth]{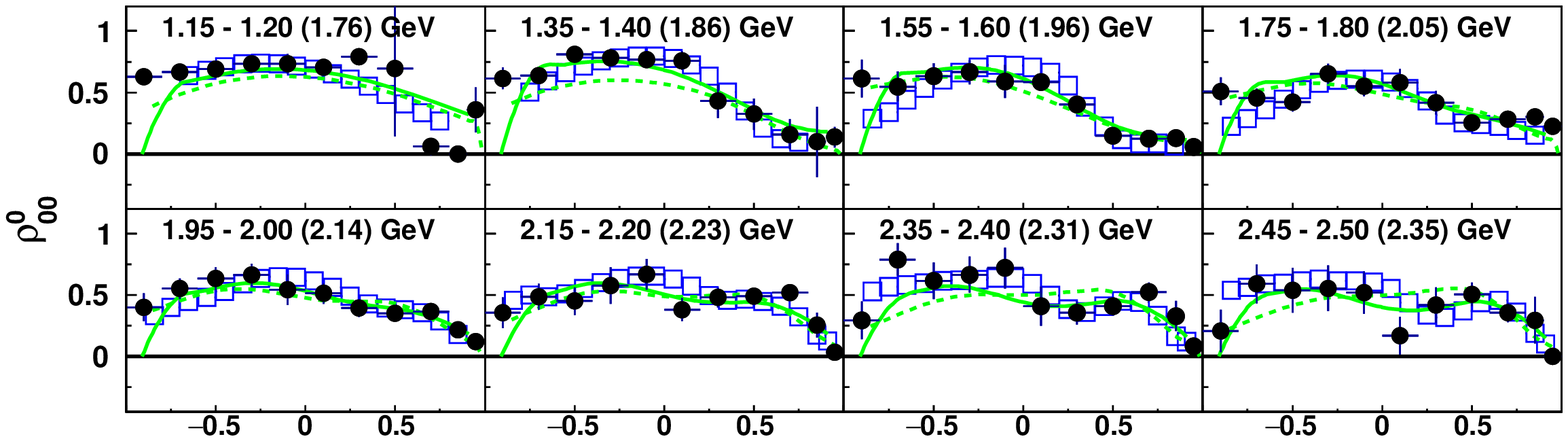}
 \includegraphics[width=0.89\textwidth]{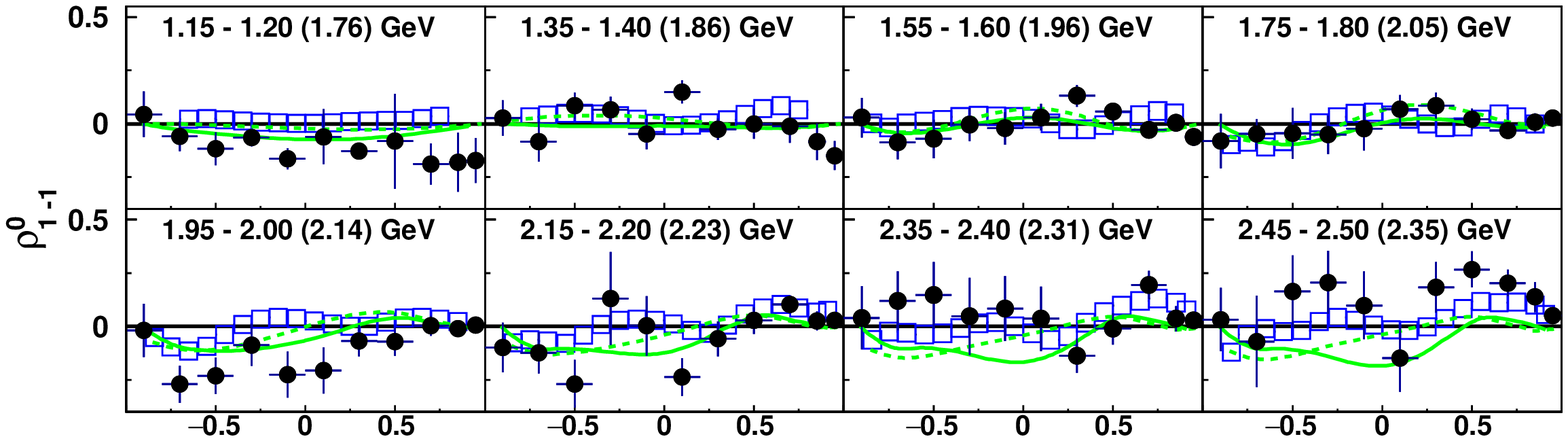}
 \includegraphics[width=0.89\textwidth]{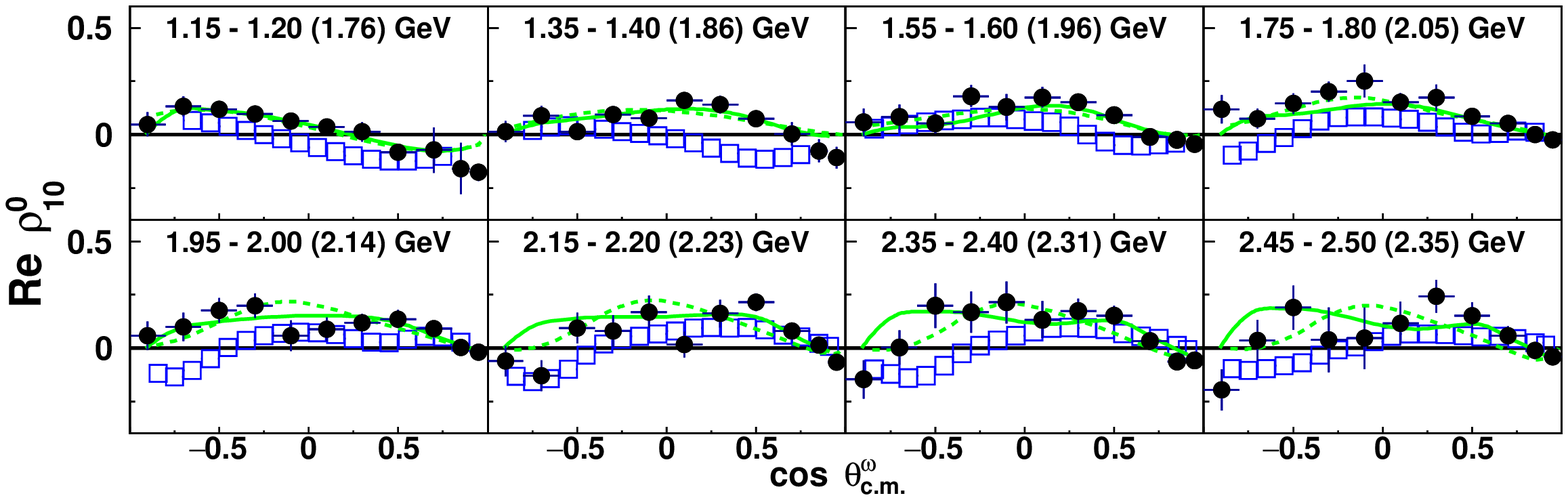}
\caption[Spin Density Matrix Elements Comparison]{(Color Online) Unpolarized spin density matrix elements 
  in the Adair frame for selected energy bins. The $\bullet$ are the SDMEs extracted in this analysis with the total
  uncertainty for each data point represented as a vertical bar.  The {\tiny \color{blue} $\square$} are the SDMEs 
  published by the CLAS collaboration~\cite{Williams:2009yj}. The Bonn-Gatchina PWA solutions are represented 
  by green lines: {\color{green}\rule[0.5ex]{0.34cm}{1pt}} (full solution), 
  {\color{green}\rule[0.5ex]{0.1cm}{1pt}\,\hspace{0.02mm}\rule[0.5ex]{0.1cm}{1pt}\,\hspace{0.02mm}\rule[0.5ex]{0.1cm}{1pt}}
  (solution without \threehalfplus{}~partial wave). Each plot is labeled with its range in initial photon energy 
 measured in the lab frame~(center-of-mass (W) energy).}
\label{fig:SDME_unpol}
\end{figure*}

\begin{figure*}[t!]
 \centering
 \includegraphics[width=0.86\textwidth]{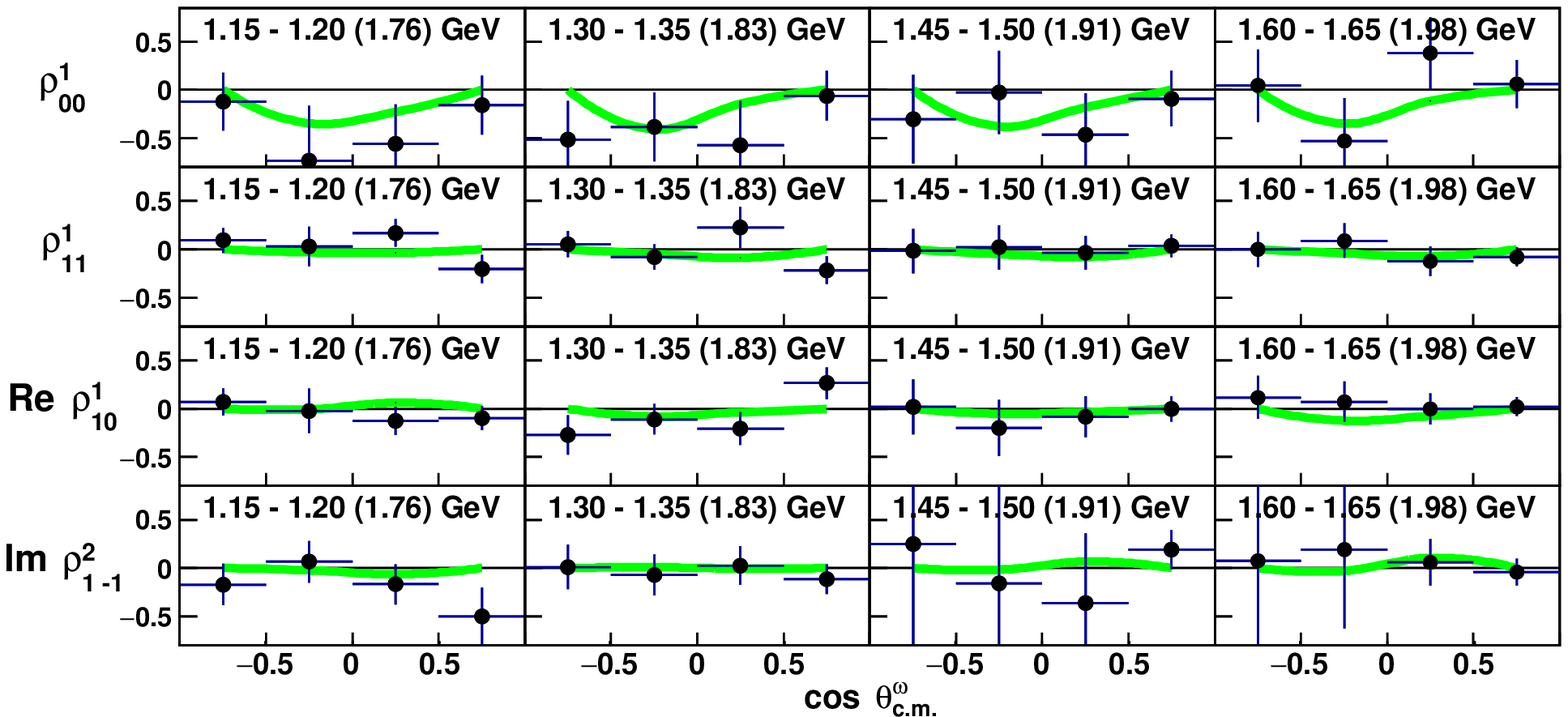}
 \caption[Spin Density Matrix Elements Comparison]{(Color Online) Polarized spin density matrix elements 
  in the Adair frame for selected energy bins. The fitted Bonn-Gatchina PWA solution is represented by a 
  {\color{green}\rule[0.5ex]{0.34cm}{1pt}} . The statistical and systematic uncertainties for each data point 
  have been added in quadrature and are represented as a vertical bar. Each plot is labeled with its range in 
  initial photon energy in the lab frame~(center-of-mass (W) energy).}
\label{fig:SDME_pol}
\end{figure*}

Figure~\ref{fig:SDME_unpol} shows representative distributions in the Adair frame, unpolarized SDMEs from this 
analysis ($\bullet$) extracted from the $\omega$ meson's radiative decay along with the data published by the 
CLAS collaboration ({\tiny \color{blue}$\square$}) in \cite{Williams:2009yj} extracted from the $\omega$ meson's 
$3\pi$ decay. These data have 11 $\cos\theta^{\,\omega}_{\rm c.m.}$ angular bins with an energy binning identical to 
the differential cross sections already shown. The comparison of these datasets for the SDME $\rho^0_{00}$ shows 
very good agreement in contrast to the normalization discrepancy in the differential cross sections. The extraction 
of SDMEs does not depend on the absolute normalization. Some differences were found in the SDME 
$\text{Re}\:\rho^0_{10}$ and to a smaller extent, mostly at higher energies, in $\rho^0_{1\:-1}$. Since the data agree
 in $\rho^0_{00}$, these differences seem to be related to the kinematic variable $\phi_d$ in Equation~\ref{Eqn:SDME0}. 
However, little impact on the results of the partial wave analysis (PWA) was observed since the $\text{Re}\:\rho^0_{10}$ 
and $\rho^0_{1\:-1}$ values are very small.

Four of the six polarized SDMEs measured in the Adair coordinate system are shown in Figure \ref{fig:SDME_pol} for 
selected energy bins. These data have four $\cos \theta^{\,\omega}_{\rm c.m.}$ bins and the same $E_\gamma$ binning as 
the above unpolarized data. With the exception of $\rho^1_{00}$, the values of the polarized SDMEs above 1.4~GeV in 
photon energy extracted in the Adair frame are very small and seem to be consistent with zero within our statistical 
uncertainties. 

\section{Partial Wave Analysis}
\label{sec:PWA}
Owing to the broad and overlapping nature of baryon resonances, amplitude analyses or PWAs need to be 
performed to extract $N^\ast$ parameters from the data. The situation is particularly complicated above the 
$\Delta$ region. Many open channels need to be considered and any reliable extraction of resonance properties 
must be based on a coupled-channel approach. While several groups have significantly contributed to our 
understanding of baryon resonances, a comprehensive PWA based on a larger database of observables has 
been performed only at very few institutions. The photoproduction data in most channels still suffer from 
an insufficient number of observables with good precision. For this reason, different groups have made 
different claims about which resonances are important and in which regions and for which processes. 
However, an improved understanding of these different solutions has been clearly observed in recent 
years due to an increasing database of high-quality measurements from current facilities. This section 
describes the results of a PWA in the framework of the Bonn-Gatchina (BnGa) coupled-channel 
approach~\cite{Anisovich:2004zz} that includes the data presented here.

The $\gamma p\to p \omega $ differential cross section has a strong peak in the forward direction; and therefore, 
a large contribution from $t$-channel exchange amplitudes is expected at higher energies. An initial fit with 
only two $t$-channel amplitudes defined by the reggezied pomeron- and $\pi$-exchange described the
forward peak fairly well but failed to reproduce the entire angular range. This fit returned very small and almost 
flat angular distributions for all three unpolarized SDMEs, and it also failed to reproduce the polarization 
observables. Contributions from nucleon resonances are a necessary additional ingredient to describe the data. 

The new data were added to the BnGa database which already included a large set of data on pion- and photo-induced 
meson production reactions, with up to two pseudoscalar mesons in the final state~\cite{Anisovich:2011fc}. In a first 
attempt at extracting resonance contributions from this analysis, additional CBELSA/TAPS $\omega$ polarization 
observables were included~\cite{Klein:2008aa,Eberhardt:2015lwa} in the PWA and all photo-nucleon and pseudoscalar 
meson-nucleon couplings from the previous BnGa solution (BG2014-02)~\cite{Sokhoyan:2015fra} were fixed. Only the 
$\omega N$~couplings of the nucleon states and $t$-channel exchange amplitudes were fitted. More than 
100 fits were tested for different initial conditions.

The elements of the density matrix are connected to one another by the production amplitudes. The SDMEs, in 
particular $\rho^0_{00}$, were therefore essential to describe the contributions from nucleon resonances. Moreover, 
the inclusion of the polarized SDMEs allowed the study of the production process in more detail and helped separate
 the natural and unnatural parity-exchange contributions. In this analysis, it was possible to distinguish between 
pomeron- (natural) and $\pi$-exchange (unnatural), which is a clear advantage over previous approaches.

Figure~\ref{fig:Dcross_EF} shows the description of the differential cross sections; the solid line represents the 
best solution. The description of the total cross section and the contribution of the five largest amplitudes for 
the best solution are shown in Figure~\ref{fig:TotCross}. The pomeron-exchange amplitude provides the largest 
contribution; it dominates the total cross section in the high-mass region. The $\pi$-exchange contributes much 
less than the pomeron-exchange. Moreover, its contribution was found to be unstable; it varied in different solutions 
from 5\,\% to 30\,\% depending on the behavior of the form factors. The description of the unpolarized SDMEs and 
polarized SDMEs is shown in Figure~\ref{fig:SDME_unpol} and Figure~\ref{fig:SDME_pol}, respectively. A discussion 
of the CBELSA/TAPS $\omega$ double-polarization observables included in this analysis and their description can 
be found in~\cite{Eberhardt:2015lwa}.

At low energies, the $J^P=$ \threehalfplus{} is the leading resonant partial wave and shows a strong peak with a 
maximum around $W = 1.8$~GeV. Such a behavior is identified with a strong $\omega N$ coupling of the 
$N(1720)$\,\threehalfplus{} state which is situated just below the reaction threshold. This state decays into the 
$\omega N$ channel with orbital angular momentum $L=1$, therefore a large contribution near threshold came as 
a surprise. The contributions from the \onehalfminus{} and \threehalfminus{} partial waves are notably smaller, in 
spite of the fact that the $\omega N$~channel can couple to these partial waves with \mbox{$L=0$}. The \threehalfplus{} 
wave has a more complex structure and hints for at least one more resonance around $W = 1.9$~GeV were found. Fitting
 without the \threehalfplus{} wave significantly deteriorated the  description of the data, particularly of the beam 
asymmetry (not shown here) and of the SDME $\rho^0_{00}$. The dashed (green) line in Figures~\ref{fig:Dcross_EF} 
and \ref{fig:SDME_unpol} shows the PWA solution without the \threehalfplus{} partial wave included.

The contribution from the \threehalfminus{} partial wave has two maxima: one at $W = 1.87$~GeV and a second 
around $2.1$~GeV. This structure is identified with the contributions from the two \threehalfminus{} states, 
$N(1875)$ and $N(2120)$. The \onehalfminus{} partial wave has a maximum close to the threshold region which is 
identified with the tail from the two lowest \onehalfminus{} states and a minimum due to the destructive interference 
with the $N(1895)$\,\onehalfminus{} state. The PWA also found a notable contribution from the \fivehalfplus{} partial 
wave. This wave has some structure close to the threshold and also around $W =2$~GeV; the latter is identified with 
the $N(2000)$\,\fivehalfplus{} state. The contributions from the $5/2^-$, $7/2^+$ and $7/2^-$ partial waves appeared 
to be smaller. In all fits, they were found to be less than 5\,\%. However, the $7/2$ partial waves play an important 
role in the description of the density matrices at masses above 2.1~GeV. They produced particular structures in the 
$\rho^0_{00}$ and $\rho^0_{1-1}$ angular distributions.

A recent CLAS PWA~\cite{Williams:2009aa} also observed a $5/2^+$ partial wave around 2~GeV and a resonance
contribution above 2.1~GeV, which was identified as $7/2^-$. The dominant $t$-channel exchange amplitude 
was found to be $\pi$-exchange; $3/2^-$ and $5/2^+$ were reported as the dominant waves at threshold. The
authors also observed a significant $3/2^+$ amplitude but did not claim resonance contributions. This partial 
wave appeared to be inconsistent with a single resonance. This is compatible with the BnGa solution. It must be
noted that the CLAS PWA was a single-channel analysis and did not include polarization observables. In the BnGa 
PWA, the multichannel approach provided signifcant constraints. The $t$-channel amplitude was identified 
by the polarized SDMEs and the additional polarization data~\cite{Klein:2008aa,Eberhardt:2015lwa} were important 
to identify clear fit minima.

In utilizing linearly-polarized photons, vector-meson photoproduction is defined by six density matrix 
elements $\rho^1_{00}$, $\rho^1_{11}$, $\text{Re}\,\rho^1_{10}$, $\rho^1_{1-1}$, $\text{Im}\,\rho^2_{10}$,
$\text{Im}\,\rho^2_{1-1}$, defined in Equations~(\ref{Eqn:SDMEPL})-(\ref{Eqn:SDMEP2}). For example, the $\omega$ 
photoproduction beam asymmetry is expressed through density matrices as~\cite{Schilling:1969um}
\be
\Sigma_\omega=\rho^1_{00}+2\rho^1_{11}\,.
\ee
Therefore, the $\Sigma$ observable alone only provides one angular distribution in contrast to the full dataset of six 
observables mentioned above.

The final fit of the full datasets with free photo- and meson-nucleon couplings did not significantly change the 
description of the $\gamma p\to p \omega $ reaction. Therefore, the present $\omega$~photoproduction data are 
compatible with previous fits to pseudoscalar meson photoproduction data. However, a good description of the new 
high-energy $p\omega$ data required a contribution from at least one new state. The available datasets do not allow to 
unambiguously determine the quantum numbers; the hypothesis of a \onehalfminus{}, $1/2^+$, $3/2^+$, or $5/2^+$ 
partial wave with a mass around 2.2~GeV leads to comparable data dscriptions. More polarization data are needed at these 
higher energies to define this state. 

In the $\cos \theta_{\rm c.m.}^{\,\omega} = [0.967-1.000]$ bin in Figure~\ref{fig:Dcross_EF}, a small discrepancy between 
the data and the best solution can be seen. It is assumed that the deviation is caused by the $t$-channel exchange of a 
higher-mass meson. It should not greatly affect the identification of contributing resonances.  

\section{Summary}
\label{sec:Summary}
The $\gamma p \to p \omega$ differential cross sections and spin-density matrix elements detected in the $\omega$
meson's radiative decay measured at the CBELSA/TAPS experiment have been presented.  The experimentally measured 
events were obtained by irradiating a liquid hydrogen target with tagged photons ranging in energy from threshold up to 
2.5 GeV. The measured cross sections presented here seem to be systematically higher than some previous measurements. 
This indicates a normalization discrepancy in particular between CBELSA/TAPS and CLAS. The spin-density matrix element, 
$\rho^0_{00}$, agrees well with the CLAS measurement and further suggests that the cross section discrepancy is related to 
an unknown issue with the absolute normalization, either at CLAS or CBELSA/TAPS.

The BnGa PWA solution indicated that the dominant contributions to the cross section near threshold were the
sub-threshold $N(1720)$\,\threehalfplus{} resonance as well as the \threehalfminus{} and \fivehalfplus{} partial 
waves. Toward higher energies, the $t$-channel contributions increased in strength. They were defined by a dominant 
pomeron-exchange and a smaller $\pi$-exchange. In addition to the $t$-channel amplitude, further contributions from 
nucleon resonances were required to describe the data. The \onehalfminus{}, \threehalfminus{}, and \fivehalfplus{} partial 
waves showed significant contributions to the PWA solution. The \onehalfminus{} wave was defined by two sub-threshold 
resonances interfering destructively with the $N(1895)$\,\onehalfminus{}  resonance. The  $N(1875)$ and  $N(2120)$ 
resonances created the two-peak structure seen in the \onehalfminus{} partial wave in Figure~\ref{fig:TotCross}. The 
\fivehalfplus{} partial wave had sub-threshold contributions along with the $N(2000)$ resonance. In addition, at least 
one previously unseen higher-mass resonance above 2~GeV was needed to describe the data.

\subsection*{Acknowledgements}
We thank the technical staff at ELSA and at all the participating institutions for their contributions to the success of this
experiment and this analysis. Financial support is acknowledged from the Deutsche Forschungsgemeinschaft (DFG) within 
the SFB/TR16 and from the Schweizerischer Nationalfonds. The presented results are partly based upon work supported by
the U.S. National Science Foundation under contract 0456463 and 0754674.

\end{document}